\documentclass[a4paper]{article}
\usepackage{amsmath,epsfig}

\newcommand{\dt}{\cdot}
\newcommand{\nn}{\nonumber}

\newcommand{\ud}{d}   
\newcommand{\grad}{\nabla}

\newcommand{\delsl}{\not\!\partial}
\newcommand{\gam}{\gamma}

\newcommand{\alp}{\alpha}

\newcommand{\sig}{\sigma}

\newcommand{\half}{{\textstyle \frac{1}{2}}}

\newcommand{\bsig}{\mbox{\boldmath $\sigma$}}
\newcommand{\bL}{\mbox{\boldmath $L$}}

\begin{document}

\noindent
{\bf\Large \textsf{Fermion absorption cross section of a \\
    Schwarzschild black hole  }}  

\vspace{0.3cm}

\noindent 
Chris Doran\footnote{e-mail: \texttt{c.doran@mrao.cam.ac.uk}}, 
Anthony Lasenby\footnote{e-mail: \texttt{a.n.lasenby@mrao.cam.ac.uk}}, 
Sam Dolan\footnote{e-mail: \texttt{s.dolan@mrao.cam.ac.uk}}
and Ian Hinder\footnote{e-mail:  \texttt{i.c.hinder@maths.soton.ac.uk}}.

\vspace{0.3cm}

\noindent
Astrophysics Group, Cavendish Laboratory, Madingley Road, \\
Cambridge CB3 0HE, UK.

\vspace{0.4cm}

\begin{center}
\begin{abstract}
We study the absorption of massive spin-half particles by a small
Schwarzschild black hole by numerically solving the single-particle
Dirac equation in Painlev\'e--Gullstrand coordinates.  We calculate
the absorption cross section $\sig(E)$ for a range of gravitational
couplings $Mm / {m_P}^2$ and incident particle energies $E$. At high
couplings, where the Schwarzschild radius $R_S$ is much greater than
the wavelength $\lambda$, we find that $\sig(E)$ approaches the
classical result for a point particle. At intermediate couplings, $R_S
\sim \lambda$, we find oscillations around the classical limit whose
precise form depends on the particle mass.  These oscillations give
quantum violations of the equivalence principle. At high energies the
cross section converges on the geometric-optics value of $27 \pi R_S^2
/ 4$, and at low energies we find agreement with an approximation
derived by Unruh.  When the hole is much smaller than the particle
wavelength we confirm that the minimum possible cross section
approaches $\pi R_S^2 / 2$.
\end{abstract}

\vspace{0.2cm}

PACS numbers:  04.70.Bw, 03.80.+r, 04.62.+v, 03.65.Pm

\end{center}

\section{Introduction}
\label{sec-introduction}

It is widely accepted that general relativity and quantum mechanics
are incompatible in their current form, yet a theory reconciling the
two remains elusive. Despite this problem, it is only at the smallest
length scales ($l < l_P$), or highest energies, that we expect
substantial modification to existing theory. At low energies, away
from spacetime singularities, the propagation of quantum fields on
gravitational backgrounds is well understood (see the books by Birrell
\& Davies~\cite{bir-quant} or Chandrasekhar~\cite{cha83}).

Interest in the absorption of quantum waves by black holes was
reignited in the 1970s, following Hawking's discovery that black holes
can emit, as well as scatter and absorb,
radiation~\cite{haw75}. Hawking showed that the evaporation
rate is proportional to the total absorption cross section. More
recently, absorption cross sections (or ``grey body factors'') have
been of interest in the context of higher-dimensional string theories.

In a series of papers~\cite{san77,san78a,san78} Sanchez considered the
scattering and absorption of massless scalar particles by an
uncharged, spherically-symmetric (Schwarzschild) black hole. Using
numerical methods she showed that the total absorption cross section
(as a function of incident frequency) exhibits oscillations around the
geometric-optics limit characteristic of diffraction patterns.
Unruh~\cite{unr76} studied the absorption of massive spin-half
particles by piecing together analytic solutions to the Dirac equation
across three regions. He showed that, in the low-energy limit, the
scattering cross section for the fermion is exactly $1/8$ of that for
the scalar particle. He also derived an approximation to the total
cross section valid at low energies, which we revisit in
section~\ref{sec-results}.

In this paper we return to the massive fermion absorption problem
studied by Unruh. We employ a different coordinate system, but retain
equivalent ingoing boundary conditions at the horizon.  Instead of
using analytical approximations we numerically integrate the Dirac
equation to calculate the absorption cross section over a range of
energies and gravitational couplings. We compare our results with the
classical cross section for a massive particle, and with Unruh's
low-energy approximation.

The natural dimensionless parameter to describe the strength of
the gravitational coupling between a black hole (of mass $M$) and a
quantum particle (of mass $m$) is given by
\begin{equation}
\alpha = \frac{G M m}{\hbar c} = \frac{M m}{m_P ^ 2} = \frac{\pi
  R_S}{\lambda_C} 
\end{equation}
where $R_S$ is the Schwarzschild radius of the hole, $\lambda_C$ is
the Compton wavelength of the quantum particle, and $m_p$ is the
Planck mass. We use the symbol $\alpha$ because it has an analogous
role in gravitation to the fine-structure constant in
electromagnetism. We expect quantum effects to be important when
$\alpha \sim 1$, whereas in the high-$\alpha$ limit classical effects
should dominate.

In first-quantised theory the capture of light and matter by a black
hole is a one-way process.  The direction of time implied by this
process is not revealed in Schwarzschild coordinates, however, as
these are manifestly time-reverse symmetric and are invalid at the
horizon.  Time-asymmetric coordinates, such as Eddington--Finkelstein
coordinates, allow the continuation of the metric across the horizon
and allow us to correctly study the properties of
wavefunctions~\cite{DGL98-grav,gap,bstates}.  We then find that
ingoing states correspond precisely to those that are regular at the
horizon.

Here, rather than using Eddington--Finkelstein coordinates, we prefer
to work with the coordinates first introduced by
Painlev\'e~\cite{painl21} and Gullstrand~\cite{gulls22}.  In these
coordinates the metric becomes
\begin{equation}
\ud s^2 = \left( 1 - \frac{2M}{r} \right) d t^2 - \sqrt{\frac{8M}{r}}
dt\, dr - d r^2 - r^2 d \Omega^2 
\label{painleve}
\end{equation}
The utility of this form of the Schwarzschild solution has recently
been highlighted by Martel \& Poisson~\cite{mart01} and
others~\cite{gap}.  For black holes (as opposed to white holes) the
negative sign for the crossterm $dt\, dr$ is the correct choice, as
this guarantees that all particles fall in across the horizon in a
finite proper time. This sign is also uniquely picked out by models in
which the black hole is formed by a collapse
process~\cite{DGL98-grav}. One advantage of this system is that the
time coordinate has a natural interpretation as the proper time
measured by an observer in freefall starting from rest at infinity.

According to general relativity the classical absorption cross section
of a Schwarzschild black hole is given by
\begin{equation}
\sig_\text{abs} = \frac{\pi M^2}{2 u^4} \left( 8 u^4 + 20u^2 - 1 + (1
+ 8u^2)^{3/2} \right) 
\label{classical_cs}
\end{equation}
where $u$ is the velocity of the particle~\cite{unr76}. In accordance
with the equivalence principle, the cross section is independent of
the particle mass $m$.  We expect that the quantum cross section will
approach this value in the limit $\alpha \gg 1$ (that is, $R_S \gg
\lambda$).

We start with the radial separation of the Dirac equation in
Painlev\'e--Gullstrand coordinates.  We then study the properties of
solutions around the horizon, identifying the physical branch of
regular solutiuons.  For unbound states $E>mc^2$ we find that the
physical solutions are composed of ingoing and outgoing waves at
infinity.  By numerically finding the ratios of these waves in any
given angular mode we are able to compute the absorption spectrum.
We use natural coordinates $G=\hbar=c=1$, except in cases where
inclusion of the factors adds clarity.

\section{The Dirac equation}
\label{sec-dirac-equation}

We let $\{\gam_0,\gam_1,\gam_2,\gam_3\}$ denote the gamma matrices in the
Dirac--Pauli representation, and introduce spherical polar coordinates
$(r,\theta,\phi)$.  From these we define the unit polar matrices
\begin{align}
\gam_r &=  \sin\!\theta (\cos\!\phi \, \gam_1 + \sin\!\phi \,
\gam_2) + \cos\!\theta\, \gam_3  \nn \\
\gam_\theta &= \cos\!\theta (\cos\!\phi \, \gam_1 + \sin\!\phi \,
\gam_2) - \sin\!\theta\, \gam_3 \nn \\
\gam_\phi &= - \sin\!\phi \, \gam_1 + \cos\!\phi \,
\gam_2.
\end{align}
In terms of these we define four position-dependent matrices $\{g_t,
g_r, g_\theta, g_\phi \}$ by
\begin{align}
\qquad
g_t &= \gam_0 + \sqrt{\frac{2M}{r}} \gam_r &
g_\theta &= r \gam_\theta \nn \\
g_r &= \gam_r &
g_\phi &= r \sin\!\theta \gam_{\phi} .
\qquad
\end{align}
These satisfy the anti-commutation relations
\begin{equation}
\{ g_\mu, g_\nu \} = 2 g_{\mu\nu} I \nn 
\end{equation}
where $g_{\mu\nu}$ is the Painlev\'e--Gullstrand metric of
equation~(\ref{painleve}).  The reciprocal matrices $\{g^t, g^r,
g^\theta, g^\phi \}$ are defined by the equation
\begin{equation}
\{ g^\mu, g_\nu \} = 2 \delta^\mu_\nu I,
\end{equation}
and both sets are well-defined everywhere except at the origin.

The Dirac equation for a spin-half particle of mass $m$ is
\begin{equation}
i g^\mu \grad_\mu \psi = m \psi,
\end{equation}
where 
\begin{equation}
\grad_\mu \psi = (\partial_\mu + \frac{i}{2} \Gamma^{\alpha \beta}_{\mu}
\Sigma_{\alpha \beta}) \psi,
\qquad
\Sigma_{\alpha \beta} = \frac{i}{4}[\gamma_\alpha, \gamma_\beta].
\end{equation}
The components of the spin connection are found in the standard
way~\cite{nak-geom} and are particularly simple in the
Painlev\'e--Gullstrand gauge~\cite{bstates},
\begin{equation}
g^\mu  \frac{i}{2} \Gamma^{\alpha \beta}_{\mu}
\Sigma_{\alpha \beta} = -\frac{3}{4r} \sqrt{\frac{2M}{r}} \gam_0.
\end{equation}
An advantage of our choice of metric is that the Dirac equation can
now be written in a manifestly Hamiltonian form
\begin{equation}
i \!\! \delsl \psi - i \gamma^0 \left(\frac{2M}{r} \right)^{1/2}
\left( \frac{\partial}{\partial r}  + \frac{3}{4r} \right) \psi = m \psi,
\end{equation}
where ${\not\!\partial}$ is the Dirac operator in flat Minkowski
spacetime. The interaction term is non-Hermitian, as the singularity
acts as a sink for probability density, making absorption possible.

The Dirac equation is clearly seperable in time, so has solutions that
go as $\exp(-iEt)$. The energy $E$ conjugate to time-translation is
independent of the chosen coordinate system, and has a physical
definition in terms of the Killing time~\cite{bstates}. We can further
exploit the spherical symmetry to seperate the spinor into
\begin{equation}
\psi = \frac{e^{-iEt}}{r} 
\begin{pmatrix}
u_1(r) \chi_\kappa^\mu (\theta, \phi) \\
u_2(r) \sigma_r \chi_\kappa^\mu (\theta, \phi) 
\end{pmatrix}
\label{trispn}
\end{equation}
where 
\begin{equation}
\sigma_r = \sin\!\theta (\cos\!\phi \, \sigma_1 + \sin\!\phi \,
\sigma_2) + \cos\!\theta\, \sigma_3.
\end{equation}
The angular eigenmodes are labeled by $\kappa$, which is a positive or
negative nonzero integer, and $\mu$, which is the total angular
momentum in the $\theta=0$ direction.  Our convention for these
eigenmodes is that
\begin{equation}
(\bsig \dt \bL + \hbar)  \chi_\kappa^\mu =  \kappa \hbar
\chi_\kappa^\mu, \quad \kappa = \pm(j+1/2) = \ldots, -2, -1, 1,2, \ldots.
\end{equation}
The positive and negative $\kappa$ modes are related by
\begin{equation}
\sigma_r  \chi_\kappa^\mu =  \chi_{-\kappa}^\mu 
\end{equation}
and are normalised so that
\begin{equation}
\int \ud \phi \int \ud \theta \sin \theta \text{ }
     {\chi^\mu_\kappa}(\theta, \phi)^\dagger \text{ }
     \chi^{\mu^\prime}_{\kappa^\prime}(\theta, \phi) = \delta_{\kappa
       \kappa^\prime} \delta_{\mu \mu^\prime}  .
\end{equation}

The trial function (\ref{trispn}) results in a pair of coupled
first-order equations  
\begin{multline}
\left(1 - 2M / r \right) \frac{\ud}{\ud r} 
\begin{pmatrix} u_1 \\ u_2 \end{pmatrix} = 
\begin{pmatrix} 1 & 2M / r \\ 2M / r & 1 \end{pmatrix}
 \\ 
\cdot
\begin{pmatrix} 
\kappa / r  &  i (E+m) - (2M/r)^{1/2}(4r)^{-1} \\  
i (E-m) - (2M/r)^{1/2}(4r)^{-1} &  -\kappa / r 
\end{pmatrix}
\begin{pmatrix} u_1 \\ u_2 \end{pmatrix}.
\label{matrixeqn}
\end{multline}
The equations have regular singular points at the origin and horizon,
as well as an irregular singular point at $r = \infty$. As far as we
are aware, the special function theory required to deal with such
equations has not been developed.  Instead we use series solutions
around the singular points as initial data for a numerical integration
scheme.

\section{Series Solutions and Boundary Conditions}

As is clear from (\ref{matrixeqn}), there is a regular singular point
in the coupled equations at the horizon, $r = 2M$. We look for series
solutions
\begin{equation}
U =
\begin{pmatrix} u_1 \\ u_2 \end{pmatrix} 
= (r-2M)^s \sum_{k=0} \begin{pmatrix} a_k \\ b_k \end{pmatrix}
(r-2M)^k  
\end{equation}
where $s$ is the lowest power in the series, and $a_k$, $b_k$ as
coefficients to be determined. On substituting into (\ref{matrixeqn})
and setting $r = 2M$ we obtain an eigenvalue equation for $s$, which
has solutions
\begin{equation}
s = 0 \hspace{0.5cm}\text{ and }\hspace{0.5cm}s = -\half + 4 i M E. 
\end{equation}
The regular root $s=0$ ensures that we can construct solutions which
are finite and continuous at the horizon. We will see later that
regular solutions automatically have an ingoing current at the
horizon. The singular branch gives rise to discontinuous,
unnormalisable solutions with an outgoing current at the
horizon~\cite{DGL98-grav}.  We therefore restrict attention to the
regular, physically-admissable solutions. The eigenvector for the
regular solution has
\begin{equation}
\begin{pmatrix} a_0 \\ b_0 \end{pmatrix} = 
\begin{pmatrix} \kappa - 2 i M (E+m) + 1/4  \\ \kappa + 2 i M (E-m) - 1/4 
\end{pmatrix} .
\end{equation}

In order to expand about infinity we need to take care of the
irregular singularity present there. There are two sets of solutions,
$U^\text{(out)}$ and $U^{\text{(in)}}$, which asymptotically resemble
outgoing and ingoing radial waves with additional radially-dependent
phase factors. To lowest order,
\begin{align}
U^\text{(out)} &= 
e^{i p r} e^{i (\phi_1 + \phi_2)} 
\begin{pmatrix} 1 \\ p / (E+m) \end{pmatrix} 
\nn \\
U^\text{(in)} &= 
e^{-i p r} e^{i (\phi_1 - \phi_2)} 
\begin{pmatrix} 1 \\ -p / (E+m) \end{pmatrix} 
\label{asympt1}
\end{align}
where the phase factors $\phi_1(r), \phi_2(r)$ are given by
\begin{equation}
\phi_1(r) = E \sqrt{8Mr},  \qquad  
\phi_2(r) = \frac{M}{p} \left(m^2 + 2p^2 \right) \ln(pr) 
\end{equation}
and the momentum $p$ is defined in the usual way, $p^2 = E^2 -
m^2$. The general (regular) solution as $r \rightarrow \infty$ is
a superposition of the ingoing and outgoing waves,
\begin{equation}
U(r \rightarrow \infty) = \alpha_\kappa U^{\text{(in)}} + \beta_\kappa
U^{\text{(out)}} 
\label{asympt2}
\end{equation}
for each angular mode. The magnitudes of $\alpha_\kappa$ and
$\beta_\kappa$ determine the amount of scattered and absorbed
radiation present.

\section{Absorption}

The spatial probability current is conserved for states with real
energy, $E > m$. For each angular eigenmode we obtain a conserved
Wronskian $W_\kappa$
\begin{equation}
W_\kappa = (u_1 u_2^\dagger + u_1^\dagger u_2) - \sqrt{2M/r} (u_1
u_1^\dagger + u_2 u_2^\dagger)  
\label{wronsk_def}
\end{equation}
which measures the total outward flux over a surface of radius $r$. At
the horizon
\begin{equation}
W_\kappa = - \left|u_1 - u_2 \right|^2 \propto - \left|a_0 - b_0
\right|^2  
\end{equation}
so the flux is inwards for all regular solutions.  On substituting the
asymptotic forms of equations (\ref{asympt1}) and (\ref{asympt2}) into
equation (\ref{wronsk_def}) we find an expression for the Wronskian in
the large-$r$ limit,
\begin{equation}
W_\kappa = - \frac{2p}{E+m} \left( \left| \alpha_\kappa \right|^2 -
\left| \beta_\kappa \right|^2  \right) 
\end{equation}
The coefficients $\alpha_\kappa$ and $\beta_\kappa$ can be determined
(up to an overall magnitude and phase) by matching the ingoing
(regular) solution at the horizon to the asymptotic form in the
large-$r$ limit. We choose the normalisation of each angular mode so
that $W_\kappa = -1$, and write the most general solution to the wave
equation as
\begin{equation}
\psi = \sum_{k \neq 0} g_\kappa \psi_\kappa 
\end{equation}
where $\psi_\kappa$ are spinors of the trial form with $u_1, u_2$ as
given by (\ref{asympt1}), (\ref{asympt2}), and $g_\kappa$ are complex
coefficients. The total absorbed flux is then just
\begin{equation}
W_{tot} = \sum_{\kappa \neq 0} \left| g_\kappa \right|^2 .
\label{ingoing_flux}
\end{equation}

We now employ a partial wave analysis to derive a simple expression
for the absorbed cross section. We write the asymptotic behaviour of
$\psi$ as the sum of a plane wave (propagating in the $\theta = 0$
direction) and an outgoing scattered wave,
\begin{equation}
\psi = e^{ipr \cos \theta} \Psi_1 + \frac{f(\theta)}{r} e^{ipr} \Psi_2
\end{equation}
where $\Psi_1, \Psi_2$ are constant spinors. The plane wave can be
decomposed into ingoing and outgoing radial waves in the large-$r$
limit. We equate the ingoing part of the plane wave with the ingoing
part of the asymptotic wave (\ref{asympt2}). Normalising the plane
wave to $2E$ particles per unit volume we find
\begin{equation}
g_\kappa \alpha_\kappa e^{i (\phi_1 - \phi_2) } = i (-1)^{\kappa+1}
\frac{\sqrt{4 \pi (E+m)}}{2p} \frac{\kappa}{\sqrt{\left| \kappa \right|}} 
\end{equation}
for each angular mode. 
The total absorption cross section $\sig_\text{abs}$ is the ratio of
the ingoing flux (\ref{ingoing_flux}) to the incident flux of the
plane wave ($2p$),
\begin{equation}
\sig_\text{abs} = \frac{\pi}{2 p (E-m)} \sum_{\kappa \neq 0}
\frac{|\kappa|}{|\alpha_\kappa|^2} .
\label{total_cs}
\end{equation}
At low energies, the $|\kappa| = 1$ states dominate the absorption,
but at higher energies we need to sum over a range of $\kappa$.

\section{Results}
\label{sec-results}

We determine the coefficients $\alpha_\kappa$ required for
(\ref{total_cs}) by matching the ingoing solution at the horizon to
the asymptotic form (\ref{asympt2}) at infinity. In a similar
calculation, Unruh~\cite{unr76} used analytic approximations to the
radial functions to find the leading contributions to the cross
section. Here, we use numerical integration of the wavefunction out
from the horizon to match the solutions, and compare our results with
the analytic approach.

\begin{figure}
\begin{center}
\includegraphics[width=11cm]{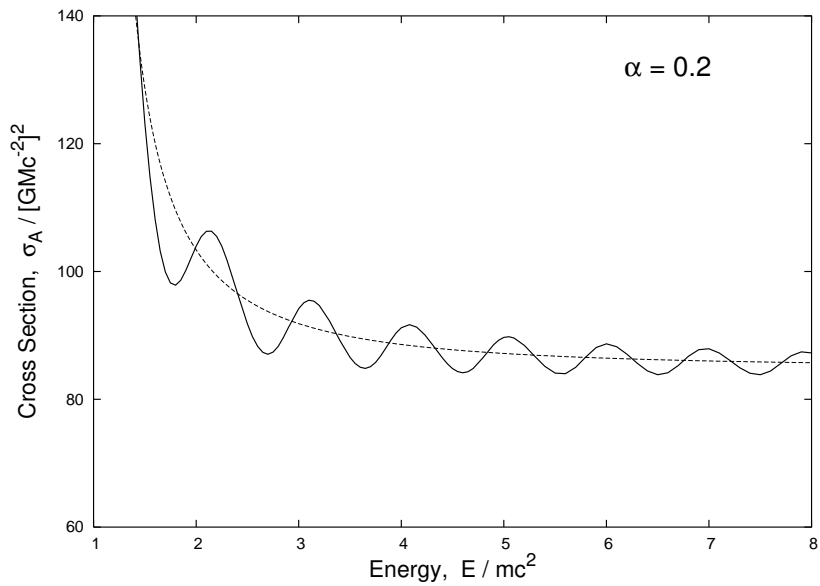}
\end{center}
\caption[dummy1]{ {\em Classical and quantum absorption cross
section}. The plot compares the absorption cross section for the Dirac
wave [solid] with the classical prediction for a point particle
[dotted], for a gravitational coupling of $\alp = Mm / {m_p}^2 =
0.2$. The cross section is plotted in units of $(GM/c^2)^2$
(proportional to the event horizon area), and the energy in units of
the rest mass energy $m c^2$.  }
\label{fig-al0pt2}
\end{figure}

Figure \ref{fig-al0pt2} compares the result of our matching
calculation at $\alpha = 0.2$ with the classical cross section of a
point particle (\ref{classical_cs}) over a range of energies. We see
that the quantum absorption cross section oscillates around the
classical value, as found by Sanchez~\cite{san77,san78a} for the
massless scalar wave. For a given $\alpha$ we find the period of these
oscillations is approximately constant, but the amplitude decays as $E
\rightarrow \infty$. As $\alpha$ is increased we find that the
magnitude and period of the oscillations decreases. In the $\alpha \gg
1$ limit we recover the classical cross section.

\begin{figure}
\begin{center}
\includegraphics[width=11cm]{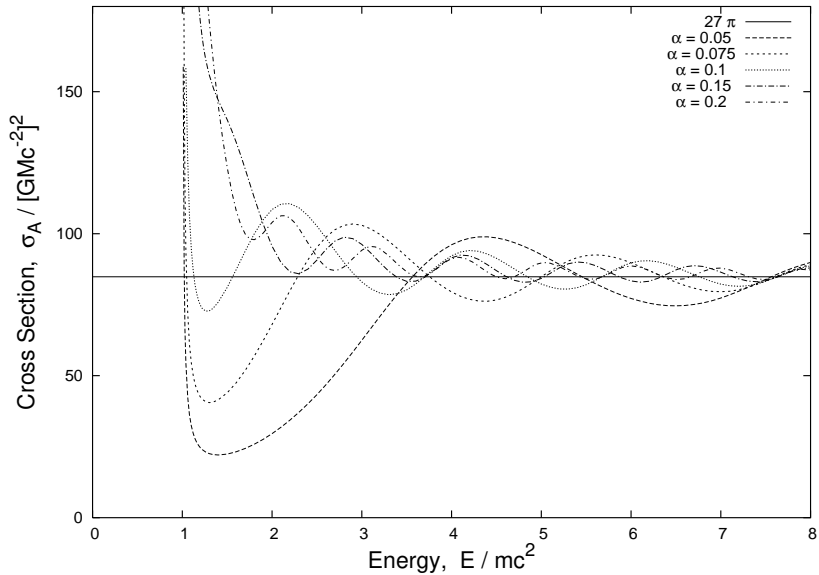}
\end{center}
\caption[dummy1]{{\em Quantum absorption cross section for a range of
couplings}.  The plot shows the cross section at a range of
gravitational couplings, $0.05 \leq Mm / {m_p}^2 \leq 0.2$. The
horizontal line is the photon limit.  }
\label{fig-al-various}
\end{figure}

Figure \ref{fig-al-various} illustrates that the precise form of the
oscillation depends on $\alpha$, and therefore on the mass of the
quantum particle. This represents a quantum-mechanical violation of
the equivalence principle. At sufficiently high energies, we see that
all cross sections tend to the photon limit of $\sig_\text{abs} = 27
\pi (GMc^{-2})^2$. As noted by Unruh, all particles travelling close
enough to light speed, $u \approx 1$, see a black hole of roughly the
same size, regardless of mass or spin.

\begin{figure}
\begin{center}
\includegraphics[width=11cm]{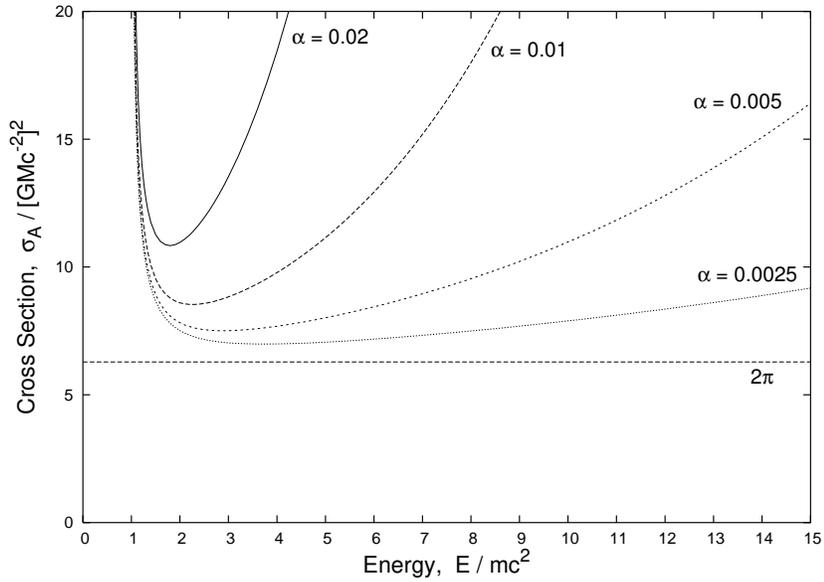}
\end{center}
\caption[dummy1]{ {\em Quantum absorption cross sections in the
$\lambda_C \gg r_S$ limit}.  The plot shows the absorption cross
section as a function of energy, for small couplings, $\alpha = Mm /
{m_p}^2 \ll 1$. In the low-energy region plotted here, the wavelength
of the Dirac particle is large compared to the black hole event
horizon. Absorption is dominated by the lowest $j = 1/2$ angular
momentum states ($\kappa = 1,-1$ states), and the low-energy
approximation of Unruh~\cite{unr76} is valid (see text). As $\alpha
\rightarrow 0$, the minimum of the cross section approaches $2 \pi$
(dotted line). In the high energy limit, all cross sections return to
the photon limit, $27 \pi$.  }
\label{fig-lowE}
\end{figure}

Unruh also showed that in the low-energy limit, the Dirac cross
section is $1/8$ of the scalar cross section, and absorption is
dominated by the lowest angular momentum modes, $|\kappa| = 1$. In
this limit he showed
\begin{equation}
\frac{\sig_\text{abs}} {(GMc^{-2})^2} \approx \frac{4 \pi^2 (1 + u^2) \alpha} {u^2 (1-u^2)^{1/2} 
\left\{ 1 - \exp( - 2 \pi \alpha (1+u^2) / u(1-u^2)^{1/2} ) \right\} }
\label{lowE_cs}.
\end{equation}
This proves to be an excellent fit to the numerical cross sections in
the low-energy regime, such as those shown in figure~\ref{fig-lowE}. A
minimum-possible cross section can be found by considering the $\alp
\sim 0$ limit of equation~(\ref{lowE_cs}), which reduces to
\begin{equation}
\frac{\sig_\text{abs}} {(GMc^{-2})^2} \approx \frac{2 \pi}{u}.
\end{equation}
The minimum value of this occurs in the $u=1$ limit, and
figure~\ref{fig-lowE} confirms that the minimum cross section
approaches $2\pi$ at low couplings.

\section{Discussion}
\label{sec-discussion}

We have shown that the absorption cross section for a Dirac wave in a
classical Schwarzschild background can be calculated by matching
ingoing solutions at the horizon to appropriate asymptotic forms at
infinity. The analysis proved particularly simple in the
Painlev\'e--Gullstrand metric, though we stress that the cross
sections calculated here do not depend on the particular choice of
gauge.  Furthermore, antiparticle solutions are be generated from
particle solutions by the transformation
\begin{equation}
(u_1, u_2, E, \kappa) \mapsto (u_1^\ast, u_2^\ast, -E^\ast, -\kappa). 
\end{equation}
It follows that the absorption cross section is invariant under charge
conjugation. 

In the large-$\alpha$ limit ($R_S \gg \lambda$) we find that the cross
section approaches the classical prediction of
equation~(\ref{classical_cs}). When $\alpha \sim 1$ ($R_S \sim
\lambda$) we observe energy-dependent oscillations about the classical
value in the $\sig$-vs-$E$ plot
(figure~\ref{fig-al0pt2}). Oscillations of this nature were previously
found by Sanchez for the massless scalar wave. The form of the
oscillations depends on $m$, so represents a quantum-mechanical
violation of the equivalence principle (figure~\ref{fig-al-various}).

In the low-energy regime, the $j=1/2$ cross section (\ref{lowE_cs})
given by Unruh is an excellent fit to our numerical results. In the
$\alpha \rightarrow 0$ ($R_S \ll \lambda$) limit we see that the
minimum-possible cross section approaches $2 \pi {(GMc^{-2})^2}$. In the
high-energy limit, all cross sections eventually converge on the photon
limit of $27 \pi {(GMc^{-2})^2}$.

\end{document}